\documentclass[%
superscriptaddress,
reprint,
 amsmath,amssymb,
 aps, physrev,
floatfix,
]{revtex4-2}
\usepackage[english]{babel}
\usepackage[version=3]{mhchem} 
\usepackage{libertine}
\usepackage{graphicx}
\usepackage{dcolumn}
\usepackage{bm}

\begin{document}


\title{Large and Versatile Plasmonic Enhancement of Photoluminescence Using Colloidal Metallic Nanocubes} 

\author{Mohammad Khaywah}
\affiliation{Université Clermont Auvergne, CNRS, SIGMA Clermont, ICCF, F-63000 Clermont-Ferrand, France}
\altaffiliation{Université Clermont Auvergne, CNRS, SIGMA Clermont, Institut Pascal, F-63000 Clermont-Ferrand, France}
\author{Audrey Potdevin}
\email{audrey.potdevin@sigma-clermont.fr}
\affiliation{Université Clermont Auvergne, CNRS, SIGMA Clermont, ICCF, F-63000 Clermont-Ferrand, France}
\author{François Réveret}
\affiliation{Université Clermont Auvergne, CNRS, SIGMA Clermont, Institut Pascal, F-63000 Clermont-Ferrand, France}
\author{Rachid Mahiou}
\affiliation{Université Clermont Auvergne, CNRS, SIGMA Clermont, ICCF, F-63000 Clermont-Ferrand, France}
\author{Youcef Ouerdane}
\affiliation{Laboratoire Hubert Curien, UMR CNRS 5516, Université Jean Monnet, Saint-Etienne, France}
\author{Anthony Désert}
\author{Stéphane Parola}
\affiliation{École Normale Supérieure de Lyon, CNRS, Univ. Lyon 1, Laboratoire de Chimie (UMR 5182), Lyon, France}
\author{Geneviève Chadeyron}
\affiliation{Université Clermont Auvergne, CNRS, SIGMA Clermont, ICCF, F-63000 Clermont-Ferrand, France}
\author{Emmanuel Centeno}
\author{Rafik Smaali}
\author{Antoine Moreau}
\email{antoine.moreau@uca.fr}
\affiliation{Université Clermont Auvergne, CNRS, SIGMA Clermont, Institut Pascal, F-63000 Clermont-Ferrand, France}

\date{\today}

\begin{abstract}
Improving phosphor photoluminescence efficiency is a key parameter to boost the performances of many optical devices. In this work, colloidal silver nanocubes,  homogeneously spread on a luminescent surface, have proved to help both injecting and extracting light in and out of the photoluminescent layer and hence contributed significantly to the enhancement of the fluorescence. This approach has been applied to two materials: the well-known Y$_3$Al$_5$O$_{12}$:Ce yellow phosphor and an optical quartz. The emission efficiency, for sol-gel derived YAG:Ce layers, has increased of 80\% in the presence of an optimal nanoparticle density -- whereas for quartz, a weakly fluorescent material, the photoluminescence signal can be enhanced by a 200-fold factor. A physical analysis based on simulations shows that the disorder is an important factor and that the surface density of Ag nanoparticles is a crucial parameter.
\end{abstract}

\maketitle

\section{Introduction}

Fluorescence enhancement is an important challenge in many optical applications ranging from lighting\cite{zhmakin_enhancement_2011} to sensing\cite{stranik_plasmonic_2005}. This is especially true for luminescent layers in LEDs for instance, whether the emitting layer is electrically or optically excited. In this context, metallic nanoparticles (MNPs) have been widely  explored to enhance the global efficiency of the process, often called the External Quantum Efficiency (EQE). MNPs are actually able to strongly interact  with light despite a typical size of a few dozen of nanometers\cite{giannini2011plasmonic} and  spreading colloidal MNPs presents the advantage of being a relatively inexpensive technique compared with lithography. In addition, MNPs offer wide diversity of geometries and sizes, leading to different optical responses\cite{wiley_maneuvering_2006,kumar_spectral_2019}. Metallic nanocubes especially have been proved to be particularly interesting in this context allowing to fabricate even complex structures\cite{moreau12b,fusella_plasmonic_2020}, while being easy to spread quite evenly on a large surface\cite{akselrod_large-area_2015}.

Metallic NPs may enhance the photoluminescence (PL) of a fluorescent layer and thus increase the EQE in two distinct ways. 

First, in the vicinity of the nanoparticles, the density of photonic states is much higher, which means an emitter is more likely to emit light if it can couple to a nanoparticle resonance\cite{anger2006enhancement}. This spontaneous emission enhancement, also called Purcell effect, is characterized by a shortening of the luminescent emitter's lifetime. The Internal Quantum Yield (IQY)
is increased in that case. However, an emitter located too close to a single MNP may excite resonances which do not radiate efficiently and light is then absorbed. Consequently, metallic nanostructures often lead to a decrease in the efficiency in the case of phosphor layers for instance\cite{lozano2016metallic}. This quenching, detrimental to the IQY, has to be avoided as much as possible for applications. However we underline that there is no universal rule for the optimal distance between the emitter and the MNP\cite{akselrod2014probing}.

Second, MNPs can help couple the guided modes supported by a luminescent layer to free space. This may increase both the Light Injection Efficiency (LIE), the efficiency with which the emitters are optically excited, and the Light Extraction Efficiency (LEE). In the case of solid-state lighting, the emitted light is hard to extract into air because there is a gap between the high refractive index of the phosphor (around 1.8 for the most common used $\ce{Y3Al5O12}$:$\ce{Ce^3+}$ (YAG:Ce) phosphor\cite{liu2010measurement}) and the air (n=1). This kind of device usually suffers from a loss of light because a large fraction (typically at least 80\% \cite{dantelle2013prepare,devys2014extraction}) of the produced photons are backscattered to the chip or trapped in dielectric structures due to Total Internal Reflection (TIR), under the form of guided modes\cite{lenef2018thermodynamics,zhmakin_enhancement_2011,leung_light_2014}. Light, in that case, can eventually escape through the edges. Allowing these guided modes to be extracted can thus significantly enhance the LEE, and ultimately the EQE.

In the past, the use of nanoparticles has allowed to enhance the photoluminescence of quantum wells (QW) optically excited and emitting in the blue region of the visible spectrum up to a factor of three\cite{henson2009plasmon,henson2010enhanced}, when all the aforementioned mechanisms play a role. The IQY in that case is actually enhanced because QW are thin enough (generally a few nanometers) to be placed in the MNPs' vicinity to leverage the Purcell effect (increase in IQY value). For thicker and electrically excited layers, like for LEDs and OLEDs, only the LEE can be increased. The enhancement in that case is sometimes difficult to assess, but is considered to be between 11\% and 40\% at most\cite{song_enhanced_2018,song_light-extraction_2009,shin_nanoparticle_2015}.

For layers that are much thicker than a few dozens of nanometers, plasmonic arrays have been recently engineered to extract the guided modes very efficiently and thus increase the LEE. The increase in the EQE is around a factor of five generally\cite{murai2017directional,kamakura2018enhanced}, but using perfectly periodical structures leads to an extremely directional emission\cite{lozano_plasmonics_2013}, which is not suitable for many applications. Disordered structures do not suffer from these drawbacks, which is another reason why the use of disordered metallic nanostructures is attractive.

Thus, there is a wide variety of plasmonic structures aimed at enhancing the luminescence of QW, LEDs or phosphor layers. The response of these devices is complex, and several phenomena are actually at play.

In this paper, we study the influence of Ag nanocubes (NCs) on the EQE of luminescent materials on which they are simply deposited. This approach allows to precisely control crucial nanostructure features such as the NCs surface density. We try first to enhance the luminescence of a YAG:Ce phosphor layer, and determine the optimal NCs density, using PL measurements upon blue excitation coupled with nanoscale microscopy study. Then we use the same approach for quartz, a material which is both very transparent and weakly luminescent, a case typically met in sensing\cite{preusser2009quartz}. Finally, our experimental results are physically interpreted using detailed simulations, providing guiding principles to reach the highest possible enhancement of the PL signal using MNPs.

\section{Methods}

We have synthesized YAG:Ce layers with well controlled surface state, deposited silver nanocubes on YAG:Ce layers and on quartz and finally characterized the structures we have obtained using electron microscopy and photoluminescence measurements. Then, we have performed two types of numerical simulations: Rigorous Coupled Wave Analysis to finely simulated the excitation of guided modes in the luminescent layers, and a Finite Difference Time-Domain (FDTD) simulations to reproduce the excitation of the luminescent layer and be able to estimate the luminescence enhancement.

{\bf Synthesis of YAG:Ce layers}.
Anhydrous yttrium chloride YCl$_3$  (99.99\% pure, ThermoFisher), and cerium chloride CeCl$_3$ (99.5\% pure, ThermoFisher), metallic potassium (98\% pure, ThermoFisher), aluminium isopropoxide (98+\% pure, ThermoFisher), acetylacetone (2,4-pentanedione, $\geq$99\%, Aldrich) and anhydrous isopropanol (iPrOH, 99.8+\% pure, Acros) were used as starting materials. 
 The sol–gel synthesis procedure used for YAG:Ce has been detailed in previous studies\cite{potdevin2009waveguiding,potdevin2007sol}. It consists in preparing separately a solution A of anhydrous yttrium and cerium chlorides dissolved in anhydrous isopropanol and a solution B of potassium isopropoxide. Solution B is slowly added to solution A under vigorous stirring: a KCl precipitate appears immediately. The mixed solution is maintained at 85°C during 1 h, then aluminium isopropoxide powder is poured directly into the solution. After further reflux for 2 h at 85°C, acetylacetone is added as chelating agent with maintaining the reflux for another 2 h at 85°C. A clear, stable and homogeneous solution (sol) is obtained together with the KCl precipitate. The latter is eliminated by centrifugation subsequent to cooling. All the synthesis steps must take place under dry argon atmosphere since alkoxides are very sensitive to moisture. The stabilized and filtered sol is then used to elaborate YAG:Ce thin films by dip-coating over well-cleaned quartz substrates (75mm*25mm*2mm). A multilayer procedure has been carried out: the substrate was slowly dipped and withdrawn into the viscous sol at a constant speed (10 cm/min). After coating, each layer was dried for several minutes at 80°C in order to evaporated volatile organic compounds and then heat-trated in a tube furnace at 400°C for 2 min to complete the elaboration process. Repeating this procedure, a 20 multicoated films were conducted in order to achieve 470 nm YAG:Ce coating.
 
 {\bf Nanocube deposition and characterization.}
 Commercial Nanocomposix NanoXact  74 nm $\pm$ 7 nm sized Ag nanocubes solutions of 2.6$\times$10$^{11}$ particles/mL were diluted to reach the desired concentration, sonicated then drop-casted over the YAG:Ce thin films  in order to produce well dispersed, aggregate free, and variable on-substrate nanocubes density. Scanning electron microscopy (SEM) images were collected at 2MAtech (Clermont-Ferrand) under high vacuum using a ZEISS Supra 55 FEG-VP instrument operating at 3kV.

{\bf Photoluminescence mapping.}
Confocal microluminescence and Raman spectra were acquired with a Labram Aramis confocal spectrometer (Horiba Jobin-Yvon) equipped with three micrometric step motors (x, y, and z directions) and a CW-He-Cd laser as a probe source. All the data were recorded using a 40x objective for the 325 nm excitation line and a CCD camera. The probe laser power was adjusted with density filters (few tens of ~µW) to avoid possible modification of the analyzed structures. The spectral mapping responses were recorded with an µm spatial resolution (at least ~ 2~µm$^2$).

{\bf Photoluminescence measurements.}
Photoluminescence measurements were performed using a continuous wave blue laser diode emitting at 450~nm. An aspheric lens was used to focus the laser beam on the sample, the spot area was equal to 650~$\mu m^2$. The signal emitted by the sample was collected through a microscope objective, focused on the slits of a 320 mm focal monochromator, and finally detected by a CCD camera.

{\bf FDTD Simulations.}
The Finite Difference Time Domain (FDTD) simulations were performed using home made codes. We considered a 4 $\mu m$ by 2 $\mu m$ cell surrounded by 1 $\mu m$ perfectly matched layers. The spatial resolution was 10 nm and the time step was about $1.7 \times 10^{-17} s$. The total calculation time was about 3.3 ps, which represents around 200 000 iterations. The averaged Poynting vector was  integrated along a segment of 4 $\mu m$ length placed 0.5 $\mu m$ above the NCs. The exaltation of the fluorescence was defined by the ratio of this averaged Poynting flux with and without the nanostructure.  
The calculation of this exaltation demands to take into account both the  electromagnetic interactions at the 450 nm excitation and at the 540 nm emission wavelengths. To model both the absorption of the pump field by the YAG:Ce layer and its fluorescence emission, we considered a four-level atomic system. The absorption of pump radiation at 450 nm wavelength promotes the electrons from level 0 to level 3. After decaying to level 2, these electrons provide the fluorescence signal at 540 nm via the transition to level 1. We described the time domain population dynamics by using a four-level rate equation formalism:
\begin{equation}
\left\{
\begin{array}{l}
  \frac{dN_3}{dt}=-\frac{N_3}{\tau_{32}}+ \frac{\mathbf{E_a}}{\Bar{h}\omega_a}  \frac{d\mathbf{P_a}}{dt} \\
  
  \frac{dN_2}{dt}=-\frac{N_2}{\tau_{21}}+ \frac{N_3}{\tau_{32}} + \frac{\mathbf{E_e}}{\Bar{h}\omega_e}  \frac{d\mathbf{P_e}}{dt} \\
  
  \frac{dN_1}{dt}=-\frac{N_1}{\tau_{10}}+ \frac{N_2}{\tau_{21}} - \frac{\mathbf{E_e}}{\Bar{h}\omega_e}  \frac{d\mathbf{P_e}}{dt} \\
  
  \frac{dN_0}{dt}=\frac{N_1}{\tau_{10}}- \frac{\mathbf{E_a}}{\Bar{h}\omega_a}  \frac{d\mathbf{P_a}}{dt} \\
\end{array}
\right.    
\end{equation}
where $N_i$ ($i=0,1,2,3$) is the population density for each of the atomic levels. The ground state population density $N_0$ was initialized with $10^{-19} \, m^{-3}$. The lifetime are $\tau_{32}=10^{-14} \,s$, $\tau_{21}=70\times 10^{-9} \,s$ and $\tau_{10}=10^{-12} \,s$ \cite{mares1987energy,he2016effects}. The angular absorption frequency $\omega_a$ is related to the atomic transition energy levels between the ground state 0 to the level 3. The angular emission frequency $\omega_e$ corresponds to the non-radiative transition from the level 2 to the level 3. The terms $\frac{\mathbf{E_a}}{\Bar{h}\omega_a}  \frac{d\mathbf{P_a}}{dt}$ and $\frac{\mathbf{E_e}}{\Bar{h}\omega_e}  \frac{d\mathbf{P_e}}{dt}$  are the induced radiation rate or excitation rate dependent on their sign. Using an electron oscillator model, the net macroscopic polarization $\mathbf{P_k}$ is induced by the applied electric field $\mathbf{E_k}$ where the the index $k$=($a$,$e$) refers to the absorption ($a$) or emission ($e$). For an isotropic medium these polarizations can be described by the following equations,
\begin{equation}
\frac{d^2\mathbf{P_k}}{dt^2}+\gamma_k \frac{d\mathbf{P_k}}{dt}+\omega_k^2 \mathbf{P_k}=\kappa_k  \Delta N_k \mathbf{E_k}
\end{equation}
where the decay rates are $\gamma_a=1/\tau_{30}$ and $\gamma_e=1/\tau_{21}$, the oscillator strength $\kappa_k= (6 \pi \epsilon_0 c_0^2 \gamma_k)/(n \omega^2_k$) with the refractive index for the YAG layer n=1.8, and  the instantaneous population density differences are  $\Delta N_a=N_0-N_3$, $\Delta N_e=N_1-N_2$. These equations are implemented through an  auxiliary differential equation (ADE) method in the FDTD algorithm\cite{al2013fdtd,alsunaidi2009general}. 

{\bf RCWA Simulations.}
The RCWA simulations\cite{granet1996efficient} were performed using home made codes using the Octave\cite{octave} platform. The field enhancement was obtained by computing a map of the temporal average of $|H_y|^2$ and $|E_y|^2$ using 161 Fourier modes inside the active layer. The field was then spatially averaged over the whole layer. This was done with and without the nanoparticles, and the ratio between the two situations was then computed, providing the field enhancement. We have checked that identical results are obtained when adding the other components of the electric and magnetic field. The dispersion curves of the guided modes were been computed relying on the freely available Moosh tools\cite{defrance2016moosh}, to find solutions of the dispersion relation for {\em a priori} complex values of the wavevector allowing to track even the leaky modes thanks to a steepest descent method.

\section{Results}

\begin{figure*}[h]
\includegraphics[width=\linewidth]{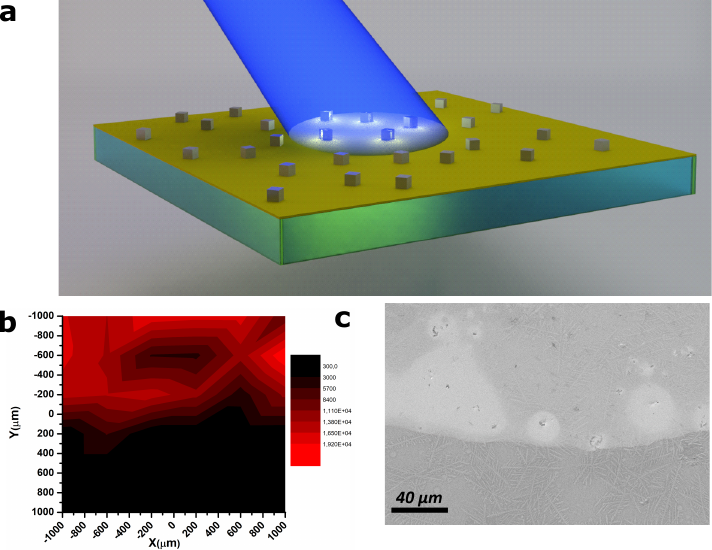}
  \caption{{\bf a} Artist view of the nanocubes on the surface of a YAG layer deposited on a quartz substrate and illuminated with a blue laser beam. {\bf b} Photoluminescence signal mapping at the edge of the area where nanocubes are deposited on a thin and inhomogeneous YAG:Ce layer. {\bf c}~Large MEB view of the edge of the zone where nanocubes have been deposited.} \label{fig:coffeering}
\end{figure*}

It is a current challenge to enhance the luminescence of a yellow phosphor layer illuminated with blue light, as this is exactly how most of the commercialized white LEDs actually work. Our first experiments have therefore been done with thin YAG:Ce layers deposited on a quartz substrate. Commercially available silver NCs with a lateral size of 75 nm were deposited on the YAG surface by drop casting. First simulations actually indicated that the optical response of NCs, especially concerning the LEE, is moderately influenced by their size. We have chosen an average size for the NCs we use. These NCs are surrounded by a 3 nm thick PVP layer\cite{moreau12b}, and are known to be electrically charged (see Supplementary Information), which generally helps to obtain uniform surface density, as schematised in Fig.\ref{fig:coffeering}a. Exploratory mappings of the luminescence using a thin YAG:Ce coating show an important but very inhomogeneous enhancement of the PL signal (see Fig. \ref{fig:coffeering}b), obviously dependent on the NCs surface density. This enhancement proved to be particularly important at the edge of the region where the NCs have been deposited (see Fig. \ref{fig:coffeering}b and c). Due to the well-known "coffee ring" phenomenon, the density of NCs is much higher there but much less homogeneous -- so that we could not determine accurately the EQE increase.

\begin{figure*}
\includegraphics[width=\linewidth]{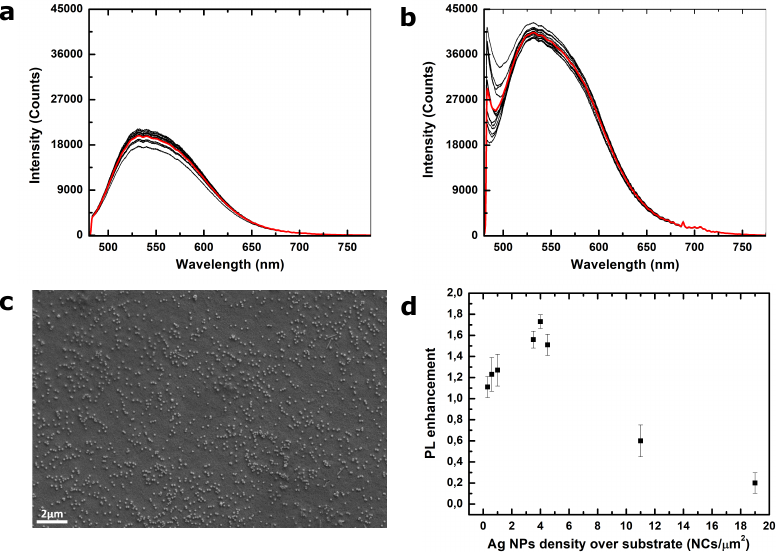}
\caption{Room-temperature emission spectra recorded under a 450 nm -excitation on a 650 µm$^2$ area in different locations, {\bf a} of the bare YAG sample and {\bf b} in the presence of nanocubes at a density of 4 NCs/µm$^2$. The average PL signal is highlighted in red while the black curves represent the PL signal at differents locations. {\bf c} Example of a SEM image of the NCs deposited on the YAG:Ce coating used to estimate the MNPs density. {\bf d} Enhancement of the photoluminescence provided by the NCs as a function of their density.}
  \label{fig:enhancement}
\end{figure*}

Preliminary simulations showed that the PL enhancement would be larger for thicker samples. As they indicated that 470 nm would be a good compromise between good optical properties and fabrication easiness, we have elaborated  homogeneous 470 nm-thick sol-gel derived YAG:Ce layers and tried to minimize the number of defects to ensure the highest PL homogeneity.
The luminescent layers were coated on a quartz substrate from a homogeneous and slightly viscous sol. A dip-coating process was used to ensure a particularly smooth surface state with reduced roughness\cite{potdevin2009waveguiding} (see Methods). 
As shown in Fig. \ref{fig:enhancement}a, the procedure leads to layers characterized by a very limited variation in the PL signal intensity and profile depending on the location considered on the sample surface (with an illumination over a 650 µm$^2$ area). Thus, we can conclude that our samples are homogeneous from a topological and optical point of view, which is absolutely necessary to assess the PL enhancement induced by the presence of metallic NCs.

After the deposition of Ag NCs, we searched for areas with a homogeneous density of metallic NCs. The structure was then illuminated using a 450 nm-excitation on a 650 µm$^2$ area. Fig. \ref{fig:enhancement} shows that the sample's PL can be increased by a maximum of 80\% for the optimal 
surface density. More precisely, Fig \ref{fig:enhancement}d clearly shows that the NCs density on the phosphor layer plays a key role in the PL enhancement. The density can be determined from SEM images recorded on the different samples (see Fig. \ref{fig:enhancement}c as an example). The maximum increase in the EQE is reached for an optimal density around 4 NCs by µm$^2$, a very reproducible result from one sample to another.  When the NCs are too concentrated, they tend to aggregate and form a metallic layer which acts partially as a mirror, reflecting more of the incoming light, and partially as an absorber, hindering light to escape. In this case, PL tends to completely disappear, whereas, when they are very sparsely distributed, the enhancement factor is close to one as could be expected when there is no NC. As explained before, measurements have been carried out in regions where NCs are homogeneously distributed which leads to a very reproducible PL enhancement presenting a variation smaller than 5 \% 
from one spot to another.

\begin{figure}
    \centering
    \includegraphics[width=0.8\linewidth]{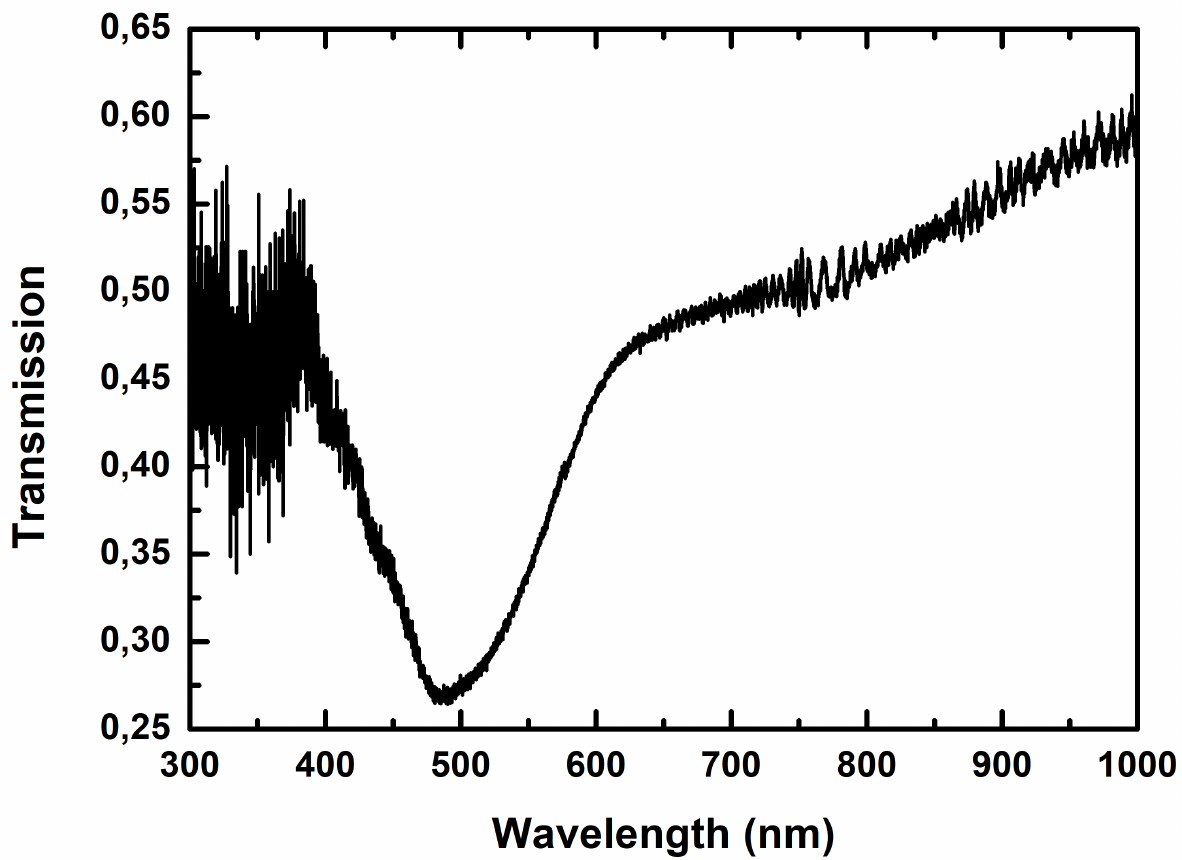}
    \caption{Transmittance spectrum of 75 nm wide Ag nanocubes deposited on a quartz substrate.}
    \label{fig:nanocubes}
\end{figure}

To better understand which mechanisms are involved in the PL enhancement, it seemed necessary to study the NCs optical properties. A way to characterize these features is to spread NCs directly on a quartz substrate and to record the transmission of the sample. This gives information on the optical response of the NCs when deposited on a substrate characterized by a refractive index higher than air (around 1.5 for quartz)\cite{nanocubes}, as is the case with YAG:Ce layers (with a refractive index of 1.8, typically\cite{liu2010measurement}). The absorption is particularly strong when plasmonic resonances are excited (between 450 and 550 nm) and the relatively narrow absorption band\cite{sherry2005localized} lying in this wavelength domain  and presented in Fig. \ref{fig:nanocubes} is consistent with the hypothesis that Ag NCs are particularly monodisperse (75 $\pm$ 10 nm, see Fig. \ref{fig:enhancement}) and not aggregated.

\begin{figure*}
\includegraphics[width=\linewidth]{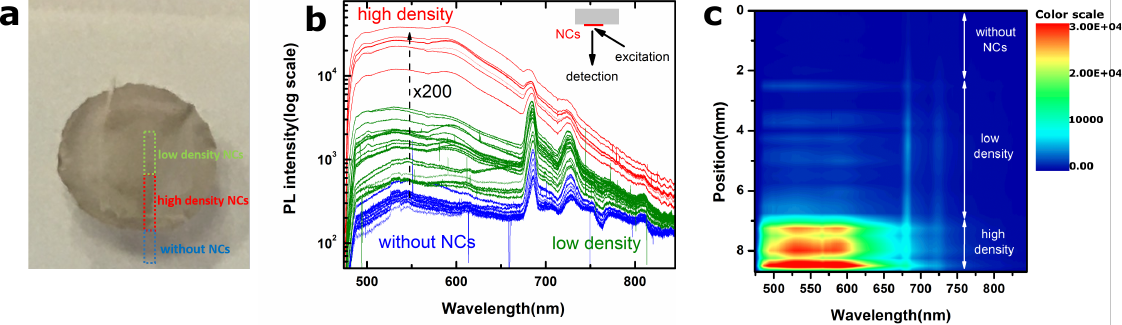}
  \caption{\textbf{a} Picture of the quartz substrate with the nanocubes (NCs) on the surface (light gray circle). The dotted lines represent the PL measurements in the region without NCs (blue), with a low NCs density (green) and with a high NCs density (red).  \textbf{b} Room-temperature emission spectra recorded along the dotted lines represented in \textbf{a}. The sample is excited by a 450~nm laser diode. In this configuration, the light emitted at the illumination spot is collected normally to the substrate, on the front. \textbf{c} PL intensity mapping as a function of the wavelength and of the illumination position showing the 200 fold factor enhancement in the most densely populated areas.}
\label{fig:edges}
\end{figure*}

During the previous study, quartz substrates have unexpectedly proved to be very weakly luminescent with a very broad emission spectrum, when excited by blue radiation. Such a luminescence is actually due to defects which are located deep inside the material structure\cite{preusser2009quartz} and a thermal treatment is usually able to make such a response disappear. This provided another platform with quite different optical properties to study the luminescence enhancement by Ag NCs. The same NCs were deposited following the same protocol as for the YAG layer described above. Resulting samples exhibit a PL characterized by an enhancement factor reaching 200 compared to that of quartz alone as shown in Fig. \ref{fig:edges}, dependent on the MNPs density. Indeed, the highest PL enhancement is obtained in areas with the highest NC density, as highlighted in Fig. \ref{fig:edges}b and c.

\begin{figure*}
\includegraphics[width=\linewidth]{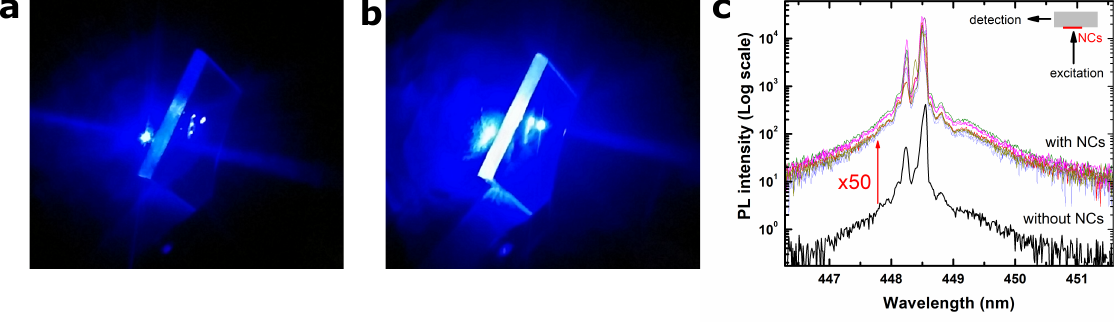}
\caption{Picture of the quartz substrate excited by the blue laser source \textbf{a} when an area without any nanocube is illuminated by the probe laser and \textbf{b} when nanocubes are actually illuminated. \textbf{c} Emission by the edges of the quartz substrate with a normal illumination using a 450 nm laser.}
  \label{fig:PLquartz}
\end{figure*}

The fact that quartz is barely absorbent and dimly luminescent allows us to better explore how such a high increase in the PL can be reached. The most obvious phenomenon which occurs when the light source illuminates the NCs deposited on the surface is the increase in the light emitted by the edges of the structure, by a factor of 50 typically, as exhibited in Fig. \ref{fig:PLquartz}. This is the sign that, thanks to the metallic NCs, guided modes are strongly excited inside the quartz layer. When there is no nanocube, the quartz being very transparent with completely flat interfaces, almost no light is converted into guided modes at excitation wavelength (Fig. \ref{fig:PLquartz}a). In the presence of NCs, the guided modes are very efficiently coupled (Fig. \ref{fig:PLquartz}b), leading to a 50 fold enhancement of the light leaking out of the slab (Fig. \ref{fig:PLquartz}c). This 50 fold increase in the LIE consequently leads to a much higher excitation of the defects present inside the material and which are responsible for the photoluminescence displayed in Fig.\ref{fig:edges}.

\section{Discussion}

The PL enhancement measured for both structures means either the LIE, the LEE or the IQY have increased. We claim that the internal quantum yield is essentially left unchanged here, just like in many similar situations\cite{lozano2013plasmonics,murai2017directional,kamakura2018enhanced}. Indeed, in order to benefit from the Purcell effect, the emitters would have to be placed in the vicinity (less than 50 nm away, typically\cite{anger2006enhancement}) from the MNPs which is essentially not the case, especially for the quartz layer where the vast majority of the emitters are at macroscopic distances from the NCs. In addition, we underline that even the YAG:Ce emission at wavelengths where the Ag NCs are clearly not resonating (typically around 600 to 700 nm) is enhanced  and the spectra present the same profile whether the PL is enhanced or not. The Purcell effect occurs when the emitted light has a frequency close to the resonance frequency of the neighbouring nanostructure, in order to benefit from the higher local density of states\cite{anger2006enhancement,nanocubes}. As all emission wavelengths are equally enhanced, this rule out completely any Purcell effect, and so any increase in the IQY. 

We recall that the efficiency of the whole process, $\eta_{EQE}$ can be written\cite{lozano_metallic_2016}
\begin{equation}
    \eta_{EQE} = \eta_{LIE} \times \eta_{IQY} \times \eta_{LEE}.
\end{equation}
The huge enhancement we have measured can be thus only be related to the increase in both LIE and LEE values. For the materials we consider here, with a refractive index ranging from 1.5 (quartz) to 1.8 (YAG:Ce), the amount of extracted light, when NCs are absent, can be estimated to be as low as 20\%, the remaining photons being trapped in the substrate or guided in the luminescent layer\cite{dantelle2013prepare}. This means that the LEE can be increased five-fold at most if all the emitted light is extracted.

Obviously, from the experiments, the LIE is largely enhanced, probably by a factor of several dozens. We can assume that the enhancement of the light emitted by the edges of the quartz slab is an evidence of this increase. In that case, the increase by a factor of 50 of the LIE and of 4 to 5 for the LEE is consistent with a total measured enhancement around 200. The whole phenomenon can be explained by the fact that even though the NCs are non-resonant at any emission wavelength, they constitute, from an optical point of view, a roughness of the surface allowing to couple guided modes both from and to air efficiently. Metallic NPs are known to present a higher scattering cross section than dielectric NPs in such a context\cite{pellegrini_light_2009}, which means very small MNPs can be as efficient to couple the guided modes as much larger and more numerous dielectric ones. Since the phenomenon is non resonant, we expect any type of metallic particle (whatever its geometry) to be able to enhance luminescence as silver nanocubes do. However the use of metals, like gold, which present a high absorption at the excitation wavelength due to interband transitions should probably lead to lower enhancement factors.

In order to better understand the influence of different parameters on the PL enhancement, we carried out two different kinds of simulations: Rigorous Coupled Wave Analysis\cite{granet1996efficient} (RCWA) and Finite Differences in Time Domain (FDTD).
Both are 2D simulations, which means invariant in the direction perpendicular to the plane of study, for simplicity reasons and because full 3D simulations of disordered NCs are extremely costly and essentially out of reach. The structures we simulated are thus infinitely long nanorods, strictly speaking. We underline that numerous studies have shown that the optical response of metallic or dielectric nanostructures consistent with the prediction of such 2D simulations, since the underlying physical mechanisms at play are identical. The resonances of NCs for instance, are well predicted by such calculations (see Supplementary Information). Bi-dimensional Mie theory allows to understand even complex features in 3D structures\cite{geints2018systematic,pitelet2019influence}. Film-coupled NCs, while they belong to a more complex kinds of resonators, are also very well described by 2D simulations and models\cite{moreau12b,lemaitre2017interferometric}.

For 2D simulations however, two polarizations can be distinguished, called $s$ (or TE) and $p$ (or TM) -- the nanostructures resonating only for the latter. 

We first used the RCWA\cite{granet1996efficient} in order to compute the electromagnetic field enhancement provided by nanorods in a 470 nm-thick YAG:Ce layer deposited on a quartz substrate, as in the above experiments. This enhancement is defined as the ratio of the electromagnetic field mean value when nanorods are present over the mean field for the multilayered structure without any metallic nanostructure\cite{defrance2016moosh}. This shows not only how LIE can be increased, but how LEE can be increased too, because of reciprocity: whenever a guided mode can be excited by an incoming plane wave, the guided mode can be efficiently coupled back to the plane wave. A large field enhancement is thus the sign of a strong coupling between free space (air) and the guided modes, providing a boost to both the LIE and the LEE.

Now we focus on the conditions under which a guided mode can be efficiently excited by the nanostructures. A guided mode is characterized by a certain wavevector $k_x$ along the structure, and thus a spatial periodicity $\lambda_x=\frac{2\pi}{k_x}$. This length is called the "effective wavelength" of the guided mode. A single NC is perfectly able to excite any guided mode with any wavevector. However, when several NCs are present the guided modes launched by the different NCs are in phase an interfere constructively only if they are 
separated by a distance which is a multiple of a guided mode effective wavelength. This constructive interference renders the excitation of the guided mode much more efficient, leading to a large enhancement of the field inside the guiding layer. This is exactly the same as saying that the guided mode wavevector has to match the wavevectors of the evanescent diffracted orders for perfectly regularly arranged nanostructures according to grating theory\cite{maystre2012diffraction}.

\begin{figure*}
\includegraphics[width=\linewidth]{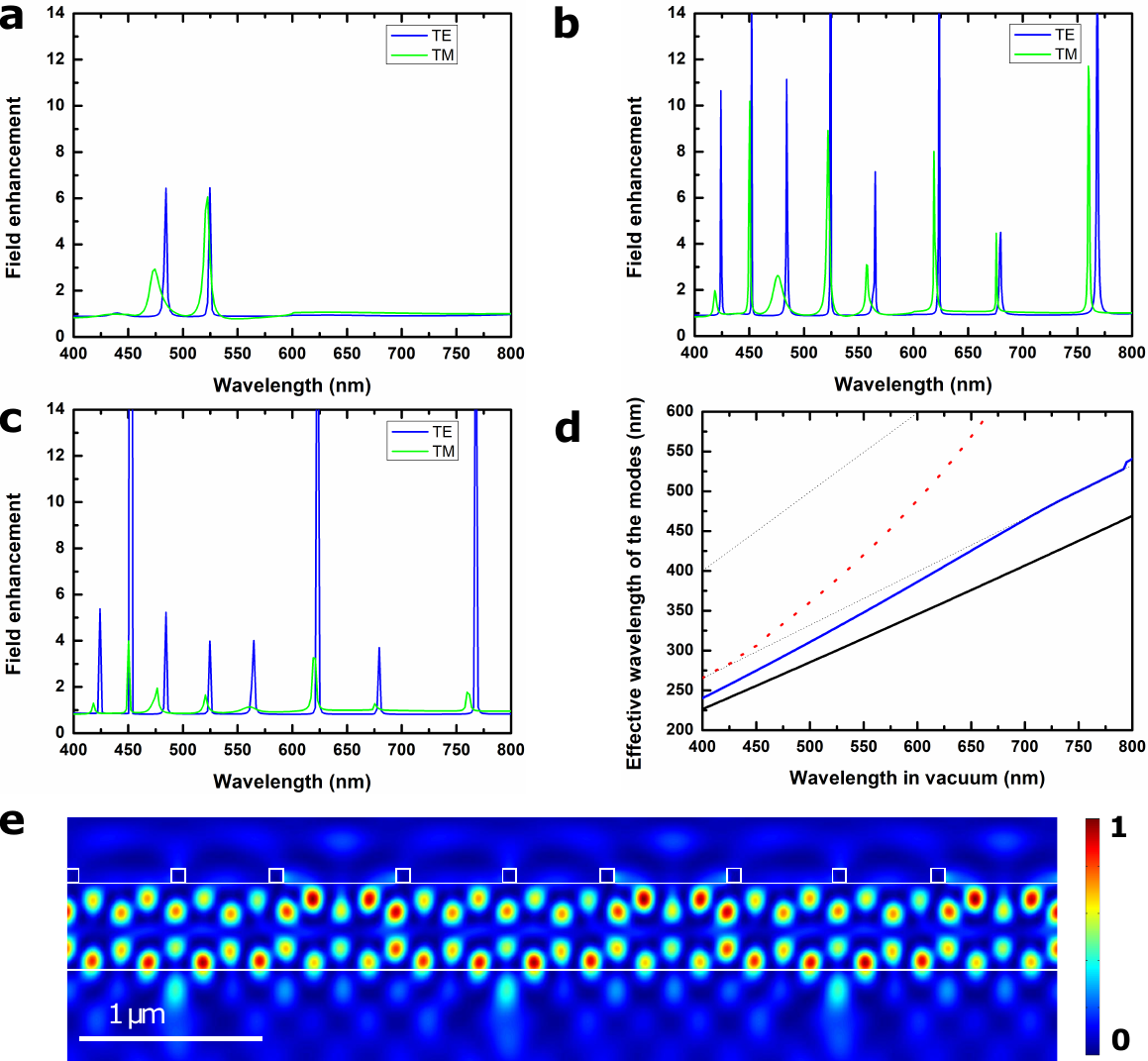}
  \caption{{\bf a} Enhancement of the electric and magnetic field when metallic nanorods are present, with a perfect periodicity of 600 nm and a normal illumination. Some guided modes are excited. {\bf b} Introducing disorder with nanocubes apart 540, 576 and 684 nm leads to a richer spectrum. {\bf c} When the nanostructures are only 180 and 306 nm apart and thus close together, some guided modes are still excited, but other are much less efficiently coupled. {\bf d} Dispersion curves in $s$ (TE) polarization for the guided (black and blue curves) and leaky modes (dashed red curve) of the 470 nm thick YAG layer. The upper (resp. lower) gray line is the air-YAG (resp. quartz-YAG) light line. {\bf e} Magnetic field for the slightly disorder structure in TM ($p$) polarization at 476 nm, corresponding to the top of a relatively broad resonance in sub-figure {\bf b}. The nanorods and the waveguide have been underlined in white.} 
  \label{fig:rcwa1}
\end{figure*}

To make this reasoning simple, we have plotted Fig. \ref{fig:rcwa1}d
the effective wavelength of the guided modes supported by a 470 nm thick YAG:Ce layer, as a function of the wavelength, for $s$ polarization here. This rather unusual way of presenting dispersion curves allows us to quickly determine, for a given distance between nanostructures, which guided mode they may excite at which frequency. Since these curves are essentially straight lines with a slope of $1/n$ typically, where $n$ is the dielectric refractive index, this suggests, as a rule of thumb, that the best typical distance between two nanostructures is the wavelength divided by the refractive index. 

In Fig. \ref{fig:rcwa1}a, the field enhancement inside the layer for the visible spectrum is presented for metallic nano-objects periodically distributed and spaced 600 nm apart. Very sharp peaks can be seen around 500 nm in both polarization, corresponding to the excitation of guided modes. More precisely, since the distance between the NCs is 600 nm, they should allow to excite guided modes with an effective wavelength of 600 nm or 300 nm (half of the distance) or 200 nm (a third) and so on. As shown in Fig. \ref{fig:rcwa1}d, the two modes supported by the luminescent layer in $s$ polarization 
present an effective wavelength $\lambda_x$ between 200 and 550 nm over the whole visible spectrum. Such modes are excited only when their effective wavelength is a fraction of the distance between two nanorods -- which means here only when it is equal to 300 nm and so only once in the whole spectrum. This leads to the two peaks in $s$ polarization which can be seen in Fig. \ref{fig:rcwa1}a and whose position can be very precisely determined  using Fig. \ref{fig:rcwa1}d.

In the simulations we  used, three nanorods can actually be placed arbitrarily in a 1800 nm range and the whole structure is then periodized, allowing us to study the impact of the nanorods' position  on the excitation of guided modes inside the YAG layer. When some disorder is introduced, by slightly changing the distance between the nanorods, a much larger number of modes can be excited. In this case, there is a variety of distances between the nano-objects, allowing them to excite the guided modes for many more different incident light wavelengths, as presented in Fig. \ref{fig:rcwa1}b. Disorder in the nanorods dispersion is therefore clearly improving the coupling efficiency between free space and the guided modes (see Fig. \ref{fig:rcwa1}e). 

When the metallic nano-objects are grouped together, Fig. \ref{fig:rcwa1}c shows that while the guided modes are actually excited, the coupling efficiency is noticeably reduced, especially in $p$ (TM) polarization. Said otherwise, clustered nano-objects are not far enough apart to contribute individually to the excitation of the guided modes and absorption may be reinforced by their proximity. In addition, when the nanorod density is really high, given the fact that the cube thickness is three times as large as the metal skin depth, the whole structure reflects a lot of light, behaving like a mirror.

This suggests that the best way to excite guided modes is to spread MNPs on the surface relatively sparsely, separated by a distance close to the average effective wavelength of the guided modes. As suggested above, a thumb rule would therefore be to consider a typical distance equal to the wavelength in vacuum divided by the refractive index of the material considered. 

The synthesis of the NCs\cite{sun2002shape} makes them electrically charged, as indicated by the measurement of their $\zeta$ potential (see Supplementary Information). This leads to a rather homogeneous spreading of NCs, which seldom aggregate compared to other nanoparticles and appear quite regularly spaced\cite{moreau12b,akselrod_probing_2014,fusella_plasmonic_2020} as previously evidenced thanks to the SEM image presented in the Fig.\ref{fig:enhancement}c . This is particularly suited to excite the highest number of guided modes as efficiently as possible. These simulations correlate well with the experimental results obtained for YAG:Ce layer and presented in Fig. \ref{fig:enhancement}. 
Actually, other kinds of metallic nanoparticles (Ag nanobeads, thinner Ag NCs) have been deposited using the same protocol, but we obtained completely different results in terms of NPs dispersion (or density) and could not observe any enhancement.  These metallic NPs were significantly aggregated; we attribute this behaviour to a much lower $\zeta$ potential that hinders NPs coulomb repulsion and auto-assembly.

The overall mechanism does not seem to be resonant. The 2D simulations we rely on allow us to consider separately the $s$ and $p$ polarizations. Yet in $s$ (or TE) polarization no surface plasmon can be excited and the nanorods are never resonant. The excitation of the guided modes seems very similar to what happens in $p$ polarization in the yellow-red part of the spectrum when the nanorods are not resonant either. In $p$ (or TM) polarization in the blue part of the spectrum, the nanorod resonances lower and broaden the peaks corresponding to the excitation of guided modes as shown Fig. \ref{fig:rcwa1}.
Nanoparticle resonances are obviously not required for the enhancement of the field inside the guiding layer. Since the phenomenon is non-resonant, we expect any type of metallic particle (whatever its geometry) to be able to enhance luminescence as silver nanocubes do. However the use of metals, like gold, which present a high absorption at the excitation wavelength due to interband transitions should probably lead to lower enhancement factors.

\begin{figure*}
    \includegraphics[width=\linewidth]{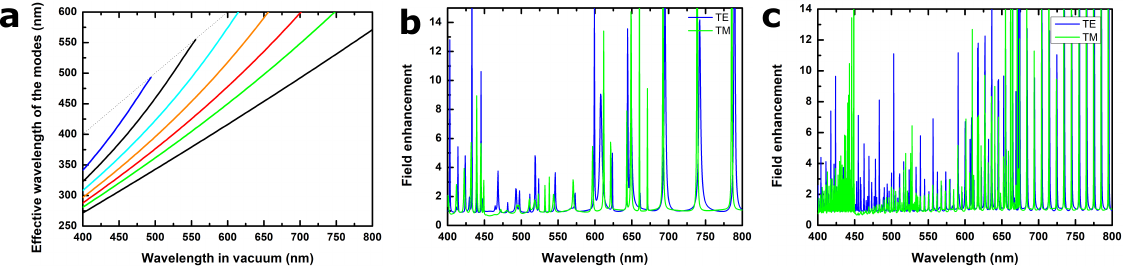}
\caption{{\bf a} Dispersion curves for the guided modes supported by a 2 µm quartz layer. {\bf b} Corresponding field enhancement with many more excitation peaks visible. The metallic nano-objects are spread as above with distances between the structures of 540, 576 and 684 nm. {\bf c} Field enhancement as a function of the wavelength in the same conditions but for a 10 µm thick layer, showing that a guided mode can be excited at almost any wavelength in the visible spectrum despite the relatively low amount of disorder.}
\label{fig:rcwa2}
\end{figure*}

We have also considered the field enhancement experienced by a quartz layer with a refractive index of 1.5 and a thickness of 2 µm on which metallic nano-objects are placed as above. A larger number of modes can be excited inside the quartz layer by the nanostructures, as illustrated by Fig. \ref{fig:rcwa2}a, leading to a very large number of resonance across the whole visible spectrum, shown in Fig. \ref{fig:rcwa2}b. This obviously suggests that the thicker the emitting layer is, the larger the number of guided modes excited and then coupled out of the structure thanks to the nano-objects. In a 10 µm thick quartz layer for instance, guided modes can be excited at almost any wavelength, explaining why the coupling of guided modes can be expected to occur for any wavelength in the visible spectrum.

We underline that such a reasoning is valid as long as the absorption of the emitting layer at the excitation wavelength can be neglected. A guided mode can be understood as an interferential reinforcement of the multiple reflections inside the guiding layer. This can occur only if the slab thickness is small compared to the penetration length of the blue light used here (see Supplementary Information).
\begin{figure}
\includegraphics[width=\linewidth]{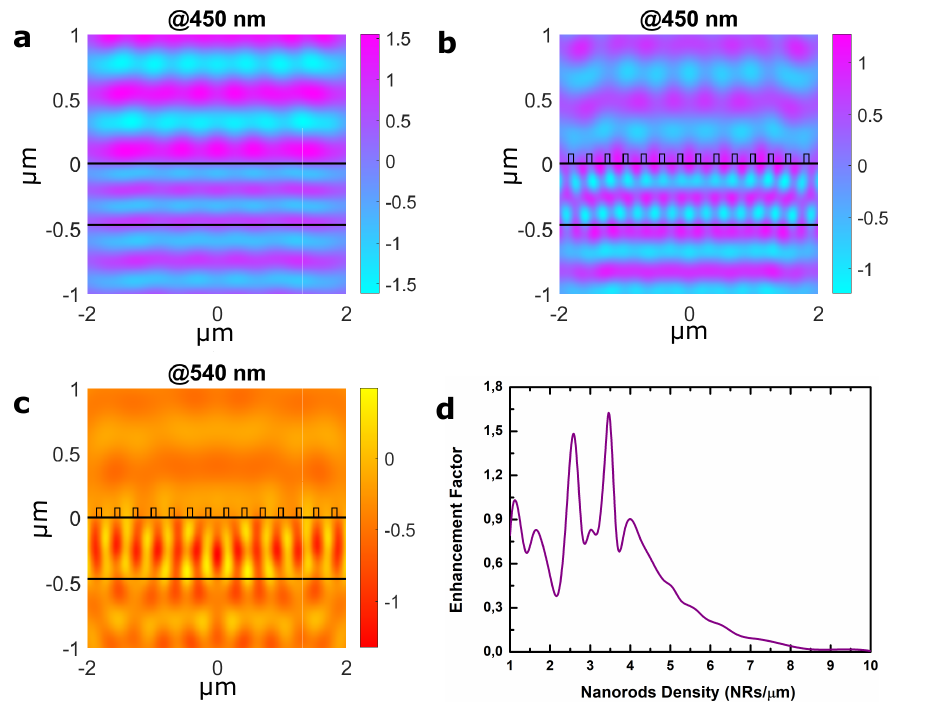}
  \caption{{\bf a} Real part of the electric field at 450 nm without the nanorods, allowing to see the wavefronts in all the materials. The map shows a standing wave above the active layer due to reflections. {\bf b} Real part of the electric field at the excitation at 450 nm with nanorods spaced 290 nm apart, showing the decrease of the standing wave, and the coupling of a guided mode inside the structure.
  {\bf c} Real part of the electric field at the emission showing the guided mode which is strongly excited by the emission. {\bf d} Enhancement emission factor provided by the nanorods as a function of the period, showing a clear maximum when both the excitation and the emission are enhanced. 
 }
  \label{fig:fdtd}
\end{figure}

To reinforce our confidence in these results, we have developed a 2D FDTD code using a four-level fluorescence model to simulate both the excitation of the emitter at 450 nm and its fluorescence emission at 540 nm. We consider the $s$ polarization case for both the excitation and emission wavelengths. In that case, the emission exaltation is therefore not related to plasmonics resonances. The 470 nm-thick YAG:Ce layer is coated on a quartz substrate. The metallic 75 nm wide nanorods are periodically set on the YAG:Ce layer on a total length equal to 4 $\mu m$. The system is excited by a plane wave in normal incidence at 450 nm. The simulations show that the nanorods enable an efficient coupling of the pump with the guided modes (Fig. \ref{fig:fdtd}a-b). At the emission wavelength (540 nm), this structure extracts a fraction of the fluorescence toward the upper half-space, as it can be seen on Fig. \ref{fig:fdtd}c. To assess the mechanism efficiency, we define the fluorescence enhancement factor as the ratio of the Poynting flux calculated in the half space containing the nanorods  with and without the metallic NCs. The fluorescence exaltation shows two maxima around 2.5 and 3.5 nanorods per µm which corresponds to the respective periods of 385 nm and 290 nm  (Fig. \ref{fig:fdtd}d). This corresponds to a trade-off here, allowing both NPs to excite guided modes at 450 nm and the guided modes excited at 540 nm to be efficiently extracted. These results strengthen the above analysis and are coherent with the experimental optimal NCs density of 4 nanoparticles per µm$^2$ evidenced in Fig. \ref{fig:enhancement}d.


\section{Conclusions}

In conclusion, we have shown that monodisperse Ag nanocubes can be used to enhance the photoluminescence of a YAG:Ce luminescent layer by 80 \%, provided the surface density of NCs is well chosen and well controlled. Furthermore, we have shown that the same technique allows to reach up to a 200-fold enhancement of the PL signal for highly transparent, weakly luminescent materials, showing how versatile such an approach can be. In the case of a quartz layer, experimental as well as theoretical results consistently point towards a large increase in the light injection and extraction efficiencies (probably by a factor of 50 and 4 respectively), to explain such a huge improvement of the external quantum efficiency. Any Purcell effect leading to an increase in the internal quantum efficiency can seemingly be excluded here. Such an enhancement can be reached with a relatively low density of Ag nanocubes, because they spontaneously spread quite regularly on the surface thanks to Coulomb repulsion, and are typically a few hundreds of nanometers apart.

Theoretical results indicate that the nanoparticles allow to efficiently couple the guided modes supported by the luminescent layer to free space (air). This is coherent with a large rise in the amount of light injected into the active material and with a large increase in the extraction efficiency too. Our simulations, carried out using two different tools, suggest that, to maximize this coupling, nanoparticles should be spread as evenly as possible, separated with a distance close to the effective wavelength of the guided modes at play. This length scale is typically close to the considered wavelength divided by the refractive index of the luminescent material, providing design principles for devices based on such an approach.

This very general technique can be used both on materials with a strong luminescence used for lighting (YAG:Ce), or on transparent samples whose PL signal is very weak but informative\cite{preusser2009quartz}. Hence, it appears as a very versatile strategy to boost photoluminescence intensity of any phosphor layer which emits in the visible range. We finally underline that while the effect demonstrated here is substantial, even better performances are definitely within our reach by optimizing the shape and density of nano-objects, as well as the thickness of the luminescent layer or the way the coating is illuminated. These improvements will be at the heart of our further works.



{\bf Acknowledgments.} This work was sponsored by a public grant overseen by the French National Agency as part of the "Investissements d'Avenir" through the IMobS3 Laboratory of Excellence (ANR-10-LABX-0016) and the IDEX-ISITE initiative CAP 20-25 (ANR-16-IDEX-0001), project Metalum. Antoine Moreau is an Academy CAP 20-25 chair holder. This work was supported by grants from the framework of the CPER DEFI MMASYF project. The authors thank the European Regional Development Fund and the Auvergne Rhône-Alpes Region funding the project. The authors would like to acknowledge the contributions of 2MAtech for the SEM imaging.

\bibliography{apssamp}

\end{document}